\documentclass[showpacs,twocolumn,floatfix,prl]{revtex4}
\usepackage{graphicx} 	
\usepackage{bm} 		
\usepackage{amssymb}
\usepackage{amsmath}
\usepackage{amsfonts}
\usepackage{epstopdf}
\usepackage{mciteplus}
\renewcommand{\section}[1]{{\par\it #1.---}\ignorespaces}

\begin{document}

\title{The importance of electron-electron interactions in the RKKY coupling in graphene}

\author{Annica M. Black-Schaffer}
 \affiliation{NORDITA, Roslagstullsbacken 23, SE-106 91 Stockholm, Sweden}

\date{\today}
\begin{abstract}
We show that the carrier-mediated exchange interaction, the so-called RKKY coupling, between two magnetic impurity moments in graphene is significantly modified in the presence of electron-electron interactions. Using the mean-field approximation of the Hubbard-$U$ model we show that the $(1+\cos(2{\bf k}_D\cdot {\bf R})$-oscillations present in the bulk for non-interacting electrons disappear and the power-law decay become more long ranged with increasing electron interactions. In zigzag graphene nanoribbons the effects are even larger with any finite $U$ rendering the long-distance RKKY coupling distance independent. Comparing our mean-field results with first-principles results we also extract a surprisingly large value of $U$ indicating that graphene is very close to an antiferromagnetic instability.
\end{abstract}


\maketitle

%
Several novel features of graphene, such as two-dimensionality, linear energy dispersion, a tunable chemical potential by gate voltage, and a high mobility have helped raising the expectation of graphene being a serious post-silicon era candidate \cite{Berger06, Avouris07,Geim07}. In this context, functionalization of graphene, especially with magnetic atoms or defects which also opens the door to spintronics \cite{Wolf01}, is of large interest. One of the most important properties of magnetic impurities is their effective interaction propagated by the conduction electrons in the host, the so-called Ruderman-Kittel-Kasuya-Yoshida (RKKY) coupling \cite{Rudermann54,*Kasuya56,*Yosida57}. This coupling is crucial for magnetic ordering of impurities but also offers access to the intrinsic magnetic properties of the host.
Several studies exist for the RKKY coupling in graphene, where both the standard perturbative approach applied to a continuum field-theoretic description of graphene \cite{Saremi07,Brey07,Bunder09} and exact diagonalization \cite{Black-Schaffer10} have been shown to give similar results. 
However, consistently, the RKKY coupling in graphene has been calculated for non-interacting electrons. This is in spite of growing evidence for the importance of electron-electron interactions in graphene with theoretical results pointing to intrinsic graphene being close to a Mott insulating state \cite{Drut09,*Drut09b,*Drut09c, Khveshchenko01b, *Khveshchenko04, Herbut06, *Herbut09b}. 
These results thus beg the question if properties such as the RKKY coupling, which are intrinsically linked to the magnetic properties of graphene, can accurately be described in a non-interacting electron picture. In this Letter, we will therefore investigate the effect of electron-electron interactions on the RKKY coupling, both in the bulk and in zigzag graphene nanoribbons (ZGNRs) where the zero-energy edge states \cite{Fujita96, *Nakada96, *Wakabayashi99} can significantly modify the RKKY behavior \cite{Bunder09, Black-Schaffer10}. We will below show that even for small to moderate strengths of the electron interactions, the RKKY coupling in the bulk is qualitatively modified and gets significantly more long-ranged than in the non-interacting electron picture. For ZGNRs the effect is even more striking as any finite $U$ causes the long-distance RKKY coupling to become distance independent. We thus conclude that it is imperative to include electron interactions when studying the RKKY coupling, and, by extension, any other properties closely related to the magnetic properties of graphene.

More specifically, we will use the mean-field approximation of the one-band Hubbard-$U$ model for graphene and include magnetic impurity spins ${\bf S} = \pm S{\bf \hat{z}}$ which couples to a graphene atom with a Kondo coupling term $J_k$:
\begin{align}
\label{eq:H}
H  =   &-t \!\! \! \! \sum_{<i,j>,\sigma} \!\!\! (c_{i \sigma}^\dagger c_{j \sigma} + {\rm H.c.}) + U \sum_{i,\sigma} \langle n_{i\sigma} \rangle n_{i-\sigma} \\ \nonumber
& + J_k\sum_{i = {\rm imp}} {\bf S}_i \cdot {\bf s}_i.
\end{align}
Here $c_{i \sigma}$ ($c_{i\sigma}^\dagger$) annihilates (creates) an electron at site $i$ with spin $\sigma$, $< \!\!i,j\!\!>$ means nearest neighbors, and ${\bf s} = \frac{1}{2}c^\dagger_\alpha {\bf \sigma}_{\alpha \beta} c_\beta$, with ${\bf \sigma}_{\alpha \beta}$ being the Pauli matrices, is the electron spin.  The constants entering, apart from $J_k$ which depends on the particular impurity moment, are the nearest neighbor hopping in graphene $t = 2.5$~eV and the on-site repulsion $U$. The value of $U$ is hard to determine exactly but, depending on the choice of exchange-correlation potential, $U/t = 1-2$ has been shown to be consistent with density functional theory (DFT) results \cite{Pisani07}. Below we are able to extract $U/t = 2.1$ when comparing the RKKY coupling in a spin chain with DFT results \cite{Pisani08}.
The expectation value of the spin-resolved electron density $n_{i \sigma} = c_{i\sigma}^\dagger c_{i\sigma}$ needs to be calculated self-consistently in Eq.~(\ref{eq:H}) and gives the spin polarization density as $s_i^z = (n_{i \uparrow} - n_{i \downarrow})/2$. The Hubbard model has been employed before in the study of graphene and ZGNRs and has been shown to yield results consistent with first-principles DFT results \cite{Fernandez07, Yazyev08}.

In standard RKKY perturbation theory \cite{Kittelbook} the leading interaction between two impurity moments at sites $i$ and $j$ is given by
\begin{align}
\label{eq:HRKKY}
H_{RKKY} =   J_{ij} {\bf S}_i \cdot {\bf S}_j,
\end{align}
with the effective RKKY coupling constant $J_{ij}$ proportional to the static spin susceptibility of the imbedding bulk. Here we will instead self-consistently solve Eq.~(\ref{eq:H})  for two impurity spins in a ferromagnetic (FM) and an antiferromagnetic (AFM) configuration, respectively and explicitly calculate the RKKY coupling as the energy difference between these two configurations: $J_{ij} = [E(FM)-E(AFM)]/2$. 
More details of our method applied to non-interacting graphene can be found in Ref.~[\onlinecite{Black-Schaffer10}].
%
%
\section{Bulk impurities}
\begin{figure}[htb]
\includegraphics[scale = 0.91]{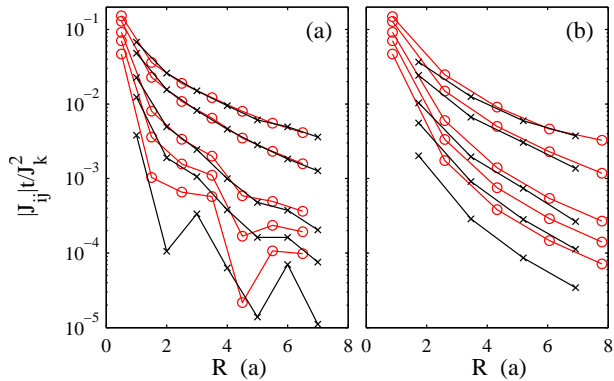}
\caption{\label{fig:bulk} (Color online) Dimensionless RKKY coupling $|J_{ij}|t/J_k^2$ as function of impurity distance $R$ in units of the lattice constant along zigzag (a) and armchair (b) directions for $U/t = 0, 1, 1.5, 2,$ and $2.15$ (increasing $|J_{ij}|$). A-A sublattice impurities (black, $\times$) has FM coupling ($J_{ij} < 0$) and A-B sublattice impurities (red, $\circ$) has AFM coupling ($J_{ij} > 0$). Lines are only guides to the eye.
}
\end{figure}

Figure~\ref{fig:bulk} shows the magnitude of the RKKY coupling as function of impurity distance $R$ along both the zigzag (a) and armchair directions (b) of the graphene lattice for several values of $U/t$. 
%
The RKKY coupling in the large $R$-limit for non-interacting graphene is $J_{ij} \propto [1+\cos(2{\bf k}_D\cdot {\bf R})]/|{\bf R}|^3$ with $J_{ij} <0$ for A-A  sublattice coupling, i.e. for impurities on the same sublattice, (black) and $J_{ij}>0$ and three times larger for A-B (or different) sublattice coupling (red) \cite{Saremi07, Black-Schaffer10}. Here ${\bf k}_D$ is the reciprocal vector for the Dirac points. Apart from minor effects due to a small $R$, these results are displayed in the lowest black and red curves in Fig.~\ref{fig:bulk}. The non-oscillatory $R$-dependence for the armchair direction is a consequence of only sampling the $\cos$-function at the graphene lattice sites.
%
When including electron interactions these results are, however, qualitatively modified even for small $U$. For $U/t = 1.5$ (middle curve) essentially all evidence of the $(1 + \cos)$-oscillations is gone as is the factor of three difference between A-A and A-B sublattice coupling. Also, the power-law decay exponent $\alpha$ changes from $3$ for $U = 0$ to around 2.3 (2.6) for $U/t = 1.5$ and 1.9 (2.1) for $U/t = 2$ for the zigzag (armchair) direction. In fact, for $U/t>2$ (uppermost curve), the armchair and zigzag RKKY couplings are equal and thus all lattice specific details have been washed out for such values of the electron interactions. With the mean-field quantum critical coupling for the AFM insulating state being $U_c/t = 2.23$ \cite{Peres04} it is perhaps not surprising that the RKKY coupling becomes independent of the small length scale details close to this point. However, what is rather unexpected is that this ``washing" out of the lattice details is clearly present even at such low values as $U/t = 1$, a value which is very likely lower than the physical value of $U$ in graphene. We thus conclude that including electron interactions is imperative when studying the RKKY interaction in graphene. Without them not only are the magnitude of the RKKY coupling grossly underestimated but, more importantly, the results do not even have a qualitatively correct $R$-dependence.

%
\section{ZGNR impurities}
\begin{figure}[htb]
\includegraphics[scale = 0.91]{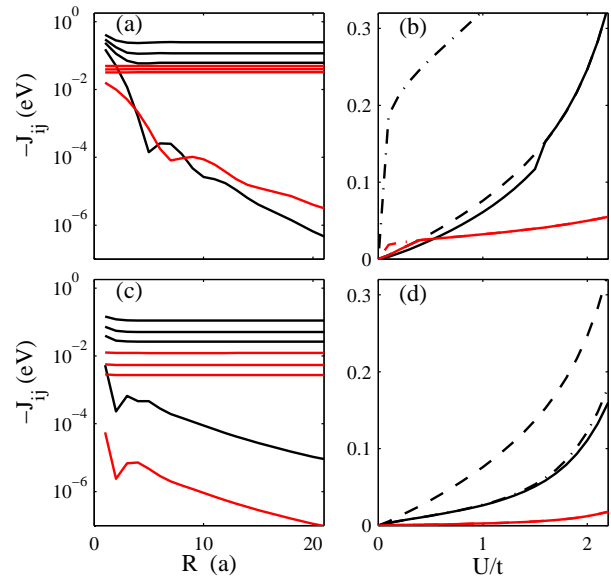}
\caption{\label{fig:ribbon} (Color online) $-J_{ij}$ for $J_k = t$ (black) and $J_k = t/10$ (red) as function of impurity distance $R$ for $U/t = 0, 1, 1.5,$ and $2$ (increasing $-J_{ij}$) (a,c) and as function of $U/t$ (b,d) for A-A edge impurities (a,b) and A-A impurities inside a narrow ZGNR of width $W = 8/\sqrt{3}a$ (c,d). Dashed lines shows $E_{DW}$ whereas dash dotted lines are equal to $J_k S s_i^z$ with $s_i^z$ being the graphene polarization at the impurity sites but with the impurities absent.
}
\end{figure}

Within the non-interacting electron picture we recently showed that for impurities along a zigzag graphene edge (A-A impurities) the RKKY interaction decays exponentially for large $R$, but that, quite counterintuitively, smaller $J_k$ gives a longer decay length \cite{Black-Schaffer10}. These results are a consequence of the extreme easiness by which an edge impurity can polarize the zero energy edge state. In contrast, for A-A impurities inside a narrow ZGNR, bulk properties of the RKKY coupling are largely regained, notably $J_{ij} \propto J_k^2/R^3$. The effect of the edge is thus only limited to edge impurities in the non-interacting limit. These results are shown in the two lowest curves in Fig.~\ref{fig:ribbon} for impurities along the edge (a) and inside the ribbon (c) for  $J_k = t$ (black) and $J_k = t/10$ (red). 
%
%
When including electron interaction this picture is dramatically changed. As is well established, any finite $U$ is going to spontaneously polarize the edge state \cite{Hikihara03, Fernandez07}, and by extension the whole ribbon \cite{Black-Schaffer10}, thus making it harder for an impurity spin to influence the polarization of the graphene. The three upper curves in (a,c) are for $U/t = 1, 1.5$ and 2, respectively. As seen, the $R$-dependence completely disappears for $R$ larger than a few unit cells for any physically relevant value of $U$ and for all impurity sites in a narrow ZGNR. The $R$-independent value of the RKKY coupling is analyzed as a function of $U/t$ in Figs.~\ref{fig:ribbon}(b,d).
As in the bulk, the FM impurity configuration is energetically favored for A-A impurities in ZGNRs, whereas the AFM configuration will require modification of the spontaneous graphene polarization to accommodate the impurity spin of the opposite orientation to said polarization. There are two $R$-independent limiting solutions for the AFM configuration of which the one with lowest energy will give an upper bound for the constant RKKY coupling. 
%
The first limiting solution has a magnetic domain wall formed between the two AFM oriented impurity spins. The magnetic domain wall formation energy per edge $E_{DW}$ is equal to the RKKY coupling for this solution and its value, calculated within Eq.~(\ref{eq:H}), is displayed with a dashed line in Figs.~\ref{fig:ribbon} (b,d). This limiting solution is not only independent of $R$ but also of $J_k$ making it especially favorable at high $J_k$-values which is also seen in Fig.~\ref{fig:ribbon}(b). 
%
For smaller $J_k$ it is, however, more likely that the most favorable AFM solution is one where the impurity spins do not noticably change the polarization of the underlying graphene, not even directly at the impurity site. The limiting AFM solution in this case is the unperturbed graphene plus the two impurities and has an energy $2J_k S s_i^z$ above that of the FM solution. Here $s_i^z$ is the graphene polarization at the site of the wrongly oriented impurity but in the absence of impurities. This unperturbed limiting solution is also naturally $R$-independent and its RKKY coupling  is displayed with dash dotted lines in Figs.~\ref{fig:ribbon}(b,d). Note that since $s_i^z$ is significantly lower inside a narrow ZGNR than on the edge, this solution yields a smaller RKKY coupling for impurities away from the edge. 
Also note that $s_i^z$ depends rather strongly on $U$, which causes both limiting solutions to increase sharply with an increasing $U$. 
%
The true RKKY coupling follow the lower of these two limiting solutions remarkably well for all four cases studied in Fig.~\ref{fig:ribbon}, including jumping form one to the other around $U = 0.3t$ for edge impurities when $J_k = t/10$. The small deviations from the unperturbed limiting solution are due to some impact of the impurity spins on the graphene polarization which locally produces small changes in $s_i^z$ in favor of a lower total energy.
The only real notable discrepancy is for edge impurities when $J_k$ is large and $U$ moderately small. Here the domain-wall limiting solution is not followed too closely but the system lowers its energy slightly by instead creating a local, half-circle shaped, domain wall around one of the wrongly oriented impurity spin. This solution naturally creates a spin-imbalance in the system as its domain wall does not propagate to other edge.
Note that both of the limiting solutions described here are always present, and thus the qualitative RKKY behavior is the same, in any system which has a spontaneous polarization in the absence of impurities, such as is the case for ZGNRs for any finite $U$. 
For edge impurities we do not expect the width of the ribbon to change the RKKY behavior as both the spontaneous edge polarization and $E_{DW}$ are weak functions of the ribbon width. However, for impurities inside a very wide ZGNR the spontaneous polarization inside the ribbon is going to be vanishingly small and bulk properties should eventually be restored for wide ribbons and small $J_k$.
We thus conclude that any finite electron interaction renders the long-distance RKKY coupling in a ZGNR $R$-independent and linearly dependent on $J_k$ for small $J_k$, but independent on $J_k$ in the limit of large $J_k$.
In addition, electron interactions make impurities inside a ZGNR behave similarly to edge impurities which is opposite to the situation for non-interacting electrons.
%
%
\section{Determining $U$}
\begin{figure}[htb]
\includegraphics[scale = 0.9]{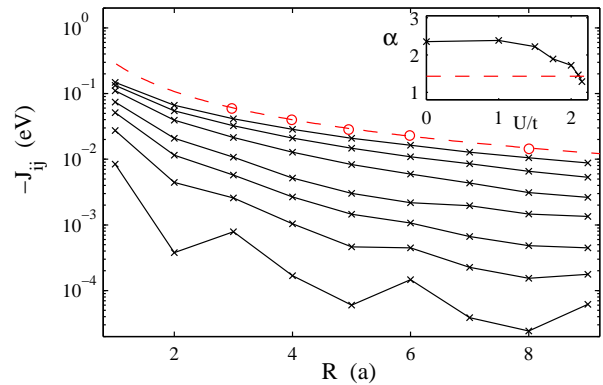}
\caption{\label{fig:dft} (Color online) $-J_{ij}$ (black, $\times$) for A-A sublattice impurities chains as function of impurity distance $R$ along the chains for $U/t = 0,1,1.5,1.75,2.1$, and $2.15$ (increasing magnitude). DFT results from Ref.~[\onlinecite{Pisani08}] (dashed red, $\circ$) are scaled with a factor 0.5. Inset shows the power law decay exponent $\alpha$ as function of $U/t$ (black, $\times$) with the exponent 1.43 from Ref.~[\onlinecite{Pisani08}] indicated with a dashed red line.
}
\end{figure}

There exist some DFT results for the RKKY coupling in graphene \cite{Pisani08,Santos09} but such studies are always limited to very small $R$ unless chains (or lattices) of impurities are studied. Fig.~\ref{fig:dft} shows the RKKY coupling for A-A sublattice impurity chains along the zigzag direction separated a distance of $25$~\AA\ as function of the impurity distance $R$ along the chains. We see that for $U = 0$ (lowest curve) characteristic non-interacting $(1+\cos)$-type oscillations are present but the chain configuration makes the RKKY coupling somewhat more long ranged than $R^{-3}$.
%
%
The oscillations however quickly disappear and the decay exponent $\alpha$ decreases (see inset) with increasing $U$. DFT results using the hybrid functional B3LYP on the same chain structure is available in Ref.~[\onlinecite{Pisani08}] and these results, scaled with an overall, unimportant prefactor, are displayed with a dashed red line in Fig.~\ref{fig:dft}. There are no oscillations in the DFT results and the exponent $\alpha$ agrees when $U/t = 2.1$ which also yields very well matched results as indicative in the main plot.
At a first glance, this might seem as a large value for the Coulomb repulsion in graphene since the mean-field AFM instability is at $U_c/t = 2.23$. However, one should keep in mind that multiple recent theoretical work have classified graphene as being very close to, if not even being, an insulator in vacuum due to strong Coulomb interactions  \cite{Drut09,*Drut09b,*Drut09c, Khveshchenko01b, *Khveshchenko04, Herbut06, *Herbut09b}.  Our results point to the fact that this state might be an AFM insulator which would be consistent with earlier results \cite{Herbut06}. 
 Our extracted value of $U$ also agrees quantitatively with earlier estimations based on the B3LYP functional \cite{Pisani07}. DFT calculations instead using the local density (LDA) or general gradient (GGA) approximations have yielded a somewhat smaller $U/t \sim 0.9-1.3$ \cite{Pisani07, EDWcomment}. It is well known that LDA suffers from electron self-interaction and therefore often underestimates $U$. B3LYP on the other hand explicitly contains an element of Fock exchange and thus tends to handle this deficiency better. This becomes especially important in strongly correlated systems but B3LYP can still reproduce the LDA results for weakly correlated materials. 
 
 With such high value of $U$ it is also natural to ask about other possible electronically driven ordered states. $U_c({\rm AFM})$ increases with doping \cite{Peres04} and thus undoped graphene is the strongest candidate for an AFM state. However, with increasing doping electronically driven $d$-wave superconductivity caused by spin-singlet nearest neighbor correlations appears for any Coulomb interaction \cite{Black-Schaffer07}. Such correlations were already proposed by Pauling and others  \cite{Paulingbook} for the $p\pi$-bonded planar organic molecules of which graphene is the infinite extension. With the Coulomb interaction extracted from the results in Fig.~\ref{fig:dft} one would need a chemical doping of $\mu = 1$~eV to reach $T_c({\rm SC}) \sim 5$~K, a value which might be achieved with, for example, chemical doping.

In summary we have shown that it is of vital importance to include electron-electron interactions when studying the RKKY coupling in graphene. Even relatively weak electron interactions qualitatively change the RKKY coupling to be significantly longer ranged and monotonically decaying in the bulk. In a ZGNR the change is even more pronounced and the $R$ dependence entirely disappears. By comparing our mean-field results we have also been able to extract a surprisingly high value for the Coulomb interactions, demonstrating that graphene might be very close to an AFM insulating instability. With such closeness to an AFM state it is rather natural that magnetic properties, such as the RKKY coupling, are going to be heavily influenced by electron interactions. 

%
%
\begin{acknowledgments}
The author thanks Sebastian Doniach, Jonas Fransson, Biplab Sanyal, Lars Nordstr\"{o}m, and Eddy Ardonne for valuable discussions.
\end{acknowledgments}

\bibliographystyle{apsrevM}

\begin{mcitethebibliography}{33}
\bibitem[{EDW()}]{EDWcomment}
\bibinfo{note}{Comparing $E_{DW}$ in Fig.~\ref{fig:ribbon} with 114 meV
  obtained within LDA \cite{Yazyev08b} we obtain $U/t = 1.3$.}\relax
\mciteBstWouldAddEndPunctfalse
\mciteSetBstMidEndSepPunct{\mcitedefaultmidpunct}
{}{\mcitedefaultseppunct}\relax
\EndOfBibitem
\end{mcitethebibliography}
\ifx\mcitethebibliography\mciteundefinedmacro
\PackageError{apsrevM.bst}{mciteplus.sty has not been loaded}
{This bibstyle requires the use of the mciteplus package.}\fi

\end{document}